\documentstyle[prb,epsfig,aps,multicol]{revtex}
\def\LJ#1{$\rm LJ_{#1}$}
\begin{document}
\draft
\title{Thermodynamics and the Global Optimization \\
of Lennard-Jones clusters}
\author{Jonathan P.~K.~Doye}
\address{FOM Institute for Atomic and Molecular Physics, 
Kruislaan 407, 1098 SJ Amsterdam, The Netherlands}
\author{David J.~Wales and Mark A. Miller}
\address{University Chemical Laboratory, Lensfield Road, Cambridge CB2 1EW, UK}
\date{\today}
\maketitle
\begin{abstract}
Theoretical design of global optimization algorithms can profitably utilize
recent statistical mechanical treatments of potential energy surfaces (PES's).
Here we analyze the basin-hopping algorithm to explain its success
in locating the global minima of Lennard-Jones (LJ) clusters, even those
such as \LJ{38} for which the PES has a multiple-funnel topography,
where trapping in local minima with different morphologies is expected.
We find that a key factor in overcoming trapping is the transformation
applied to the PES which broadens the thermodynamic transitions.
The global minimum then has a significant probability of occupation at
temperatures where the free energy barriers between funnels are surmountable.
\end{abstract}
\begin{multicols}{2}
\section{Introduction}
Global optimization is a subject of intense current interest, both scientific and
commercial. For example, improved solutions to optimization problems, such as the 
travelling-salesman problem and the routing of circuitry on a chip, could lead to
cost reductions and improved performance.\cite{KirkSA}
In the chemistry and physics communities motivation is often provided by the 
structural insight which can be derived from finding the lowest energy configuration
of a (macro)molecular system.  
Therefore, the development of better global optimization algorithms
is an important task. It is a challenge which needs to be guided by theoretical insight
rather than proceeding on purely empirical or intuitive grounds.

The main difficulty associated with global optimization
is the exponential increase in the search space with system size.
One expression of this problem is Levinthal's paradox which points out the
apparent impossibility of finding the native state of a 
protein.\cite{Levinthal}
The number of possible conformations of a typical protein
is so large that it would take longer
than the age of the universe to find the native state if
the conformations were sampled randomly, even at an unfeasibly
rapid rate.
This problem can be more rigorously defined using computational
complexity theory.\cite{Garey}
Finding the global minimum of a protein\cite{Ngo94} or
a cluster\cite{Wille} is `NP-hard', a class of problems for which 
there is no known algorithm that is guaranteed to find a solution in
polynomial time.

However, we know that proteins are able to reach their native state
relatively rapidly.
The resolution of Levinthal's paradox lies in the fact that the search
is not random---the paradox results from the implicit assumption that the
potential energy surface (PES) is flat.
In reality, the topography of the PES is crucial in determining the 
ability of a system to reach the global minimum.\cite{JD96c}
For example, the system is thermodynamically more likely (than random) 
to be in those regions of the PES that are low in energy, 
and downhill pathways can guide the system towards certain configurations. 
A funnel, a set of downhill pathways that converge on a single low energy 
minimum, combines these two effects. It has been suggested that the 
PES's of proteins are characterized by a single deep funnel and that
this feature underlies the ability of proteins to fold to 
their native state.\cite{Leopold,Bryngel95}
Indeed it is easy to design model single-funnel PES's that result in
efficient relaxation to the global minimum,
despite very large configurational spaces.\cite{JD96c,Zwanzig92,Zwanzig95}

Therefore, global optimization should be relatively easy for those systems
with a single-funnel PES topography. 
However, there are many systems for which the topography of the PES does
not permit escape from the huge number
of disordered configurations. 
These systems are more likely to form glassy or amorphous structures
on cooling. In such cases it is very difficult
for global optimization to succeed. 
This is especially true if the optimization method, 
e.g.\ simulated annealing, tries to mimic some physical process, because
the natural time scales for the formation of order are much longer than those that
can be probed by computer simulation.

It has been suggested that the PES's of such `glassy' materials are
rough, with many funnels leading to different low energy amorphous
configurations, rather than to the crystal.\cite{Still88,Still95,Rose93b} 
However, one does not need to have much complexity to make global optimization 
difficult; two funnels are enough if the system is more likely to 
enter the secondary funnel on descending the PES.\cite{JD96c}
In this paper we examine one such example from the realm of atomic clusters 
and compare it with another much easier problem.
Our aims are to pinpoint why many global optimization algorithms 
are likely to fail for this cluster, to understand the reasons for the 
success of the `basin-hopping' method and to deduce lessons for 
the design of global optimization algorithms. 
In particular, we highlight the influence of the thermodynamics of the clusters
on these questions.
A brief description of some of the results has already appeared.\cite{JD98a}

\section{Lennard-Jones clusters}

\begin{figure}
\begin{center}
\epsfig{figure=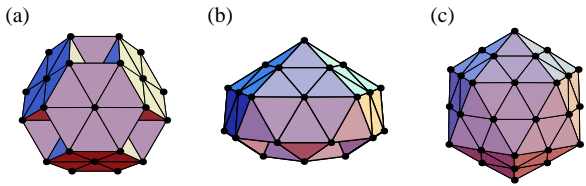,width=9.2cm}
\vglue -0.3cm
\begin{minipage}{8.5cm}
\caption{\label{piccies} (a) The LJ$_{38}$ global minimum is an fcc truncated octahedron ($E=-173.928427\epsilon$).
(b) Second lowest energy minimum of LJ$_{38}$ ($E_c=-173.252378\epsilon$). 
The structure is an incomplete Mackay icosahedron. 
(c) The LJ$_{55}$ global minimum is a Mackay icosahedron ($E_c=-279.248470\epsilon$).
}
\end{minipage}
\end{center}
\end{figure}

The Lennard-Jones (LJ) potential,\cite{LJ} which provides a reasonable description of
the interactions between rare gas atoms, is given by 
\begin{equation}
E = 4\epsilon \sum_{i<j}\left[ \left(\sigma\over r_{ij}\right)^{12} - \left(\sigma\over r_{ij}\right)^{6}\right],
\end{equation}
where $\epsilon$ is the pair well depth and $2^{1/6}\sigma$ is the 
equilibrium pair separation.
LJ clusters have been much used as a test system for global optimization methods
that are designed for configurational problems, 
and it is likely that nearly all the global minima up to 150 atoms have now been 
found.\cite{Northby87,Xue,Coleman,Pillardy,JD95c,JD95d,Deaven96,WalesD97,Leary97,Barron97,CCDLJ}
Most of the global minima have structures that are based upon the Mackay icosahedra,\cite{Mackay} 
but there are a number of exceptions.
The \LJ{38} global minimum is a face-centred cubic (fcc) truncated octahedron (Fig \ref{piccies}a), 
and for 75--77 and 102--104 atoms the global minima are based on Marks decahedra.\cite{Marks84}
These clusters are interesting because the global minima are much more difficult 
to find by an unbiased global optimization method. 
They were all initially discovered by construction using physical insight.\cite{JD95c,JD95d}
Since then the truncated octahedron has been found by a number of 
methods.\cite{Pillardy,Deaven96,WalesD97,Leary97,Niesse96a,Barron96}
The Marks decahedral global minima 
have only been found by the `basin-hopping' approach,\cite{WalesD97,Learyunpub} the method we analyse in
this paper, and a modified genetic algorithm which searches the same transformed surface.\cite{markham} 

For the above reasons we choose to examine in more detail the behaviour of \LJ{38}, and
as a contrast \LJ{55}. The \LJ{55} global minimum is a complete Mackay icosahedron\cite{Mackay} 
(Figure \ref{piccies}c), a structure that is easily found by any reasonable global 
optimization algorithm.

\begin{figure}
\begin{center}
\epsfig{figure=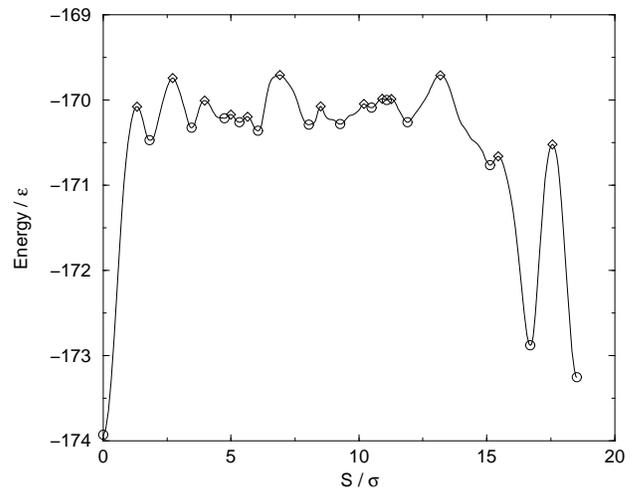,width=8.2cm}
\vglue0.1cm
\begin{minipage}{8.5cm}
\caption{\label{38path} Pathway between the fcc global minimum of LJ$_{38}$ and 
the lowest energy icosahedral minimum. The transition states on the pathway are marked by
diamonds and the minima by circles.
}
\end{minipage}
\end{center}
\end{figure}

\begin{figure}
\begin{center}
\epsfig{figure=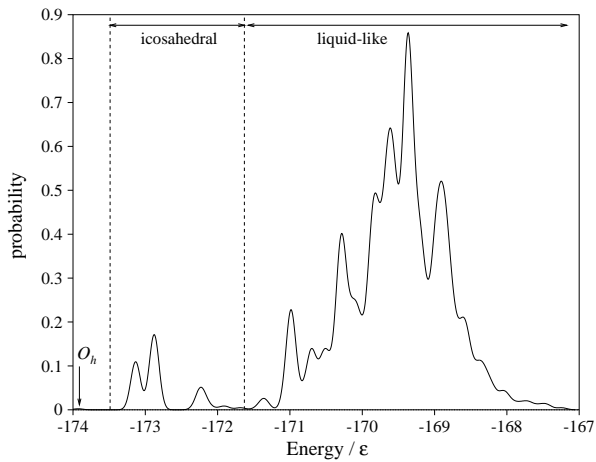,width=8.2cm}
\vglue0.1cm
\begin{minipage}{8.5cm}
\caption{\label{38.minE} Probability distribution for the potential energy of LJ$_{38}$ minima
obtained by systematic quenching from a microcanonical molecular dynamics (MD) run at $E=-148.2\epsilon$. 
866 distinct minima were obtained from the 2500 quenches. 
We have divided the minima up into three sets
by their energy, i.e.~the $O_h$ global minimum, icosahedral minima and minima that
are associated with the liquid-like state.
}
\end{minipage}
\end{center}
\end{figure}

For \LJ{38} the second lowest energy minimum is an incomplete Mackay icosahedron with $C_{5v}$ 
point group symmetry (Figure \ref{piccies}b), which lies only $0.676\epsilon$ higher in energy
than the fcc global minimum. 
As might be expected from their structural dissimilarity, the two lowest energy minima are 
well separated on the PES. 
Using a method which walks over the PES from minimum to minimum via 
transition states,\cite{JD97a,Barkema96a,Mousseau97} we have been able to find paths connecting
these two structures. 
The lowest energy path that we found is depicted in Figure \ref{38path}.\cite{pathnote}
It clearly shows that there is a large activation energy associated with passage between the two lowest
energy minima. The fcc and icosahedral regions of configuration space 
represent two distinct funnels on the PES. 
The minima at the top of the barrier are from the region of configuration space
associated with the liquid-like state (Figure \ref{38.minE}).
At temperatures below the melting point the Boltzmann weights of these states are small
implying that there is also a large free energy barrier between the two funnels.\cite{barriers}
The dynamics confirm this picture; 
below the melting point a simulation run started in one of the funnels always remains trapped 
in that funnel.

The shapes of the fcc and icosahedral funnels are also rather different.
For the fcc funnel there is a large energy gap ($2.072\epsilon$)
between the truncated octahedron and the next lowest energy minimum.
In contrast there are many low energy icosahedral structures; indeed
there are at least 24 icosahedral minima lower in energy than the second 
lowest energy fcc minimum (Figure \ref{38.minE}).
The fcc funnel is narrow, whereas the icosahedral funnel is wider and has 
a broad, fairly flat bottom.

\begin{figure}
\begin{center}
\epsfig{figure=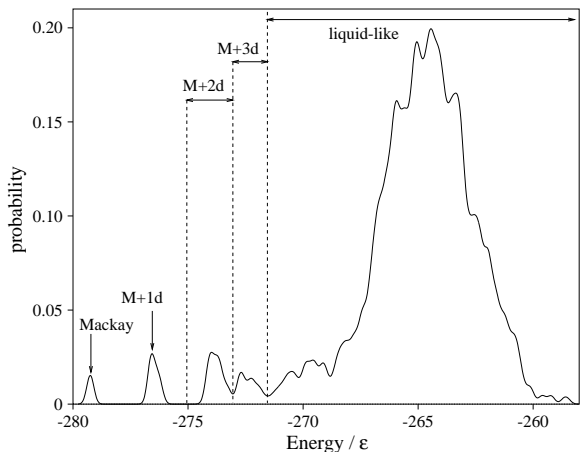,width=8.2cm}
\vglue0.1cm
\begin{minipage}{8.5cm}
\caption{\label{55.minE} Probability distribution for the potential energy of LJ$_{55}$ 
minima obtained by systematic quenching from a microcanonical MD run at $E=-209\epsilon$. 
1153 distinct minima were obtained from the 1191 quenches. 
We have divided the minima up into five sets
by their energy: the Mackay icosahedral global minimum, Mackay icosahedra with differing number of defects (M+$n$d)
and those that are associated with the liquid-like state.
}
\end{minipage}
\end{center}
\end{figure}

The \LJ{55} PES is very different. 
All the low energy minima for \LJ{55} are based on the Mackay icosahedron.
The first three peaks above the global minimum in the probability distribution of minima in Figure \ref{55.minE}
represent Mackay icosahedra with one, two and three surface defects, respectively.\cite{JD95b} 
Furthermore, the Mackay icosahedron is $2.644\epsilon$ lower than any other minimum. 
The PES of \LJ{55} has a single deep funnel.

\section{Thermodynamics of \LJ{38} and \LJ{55}}
\label{sect:thermo.un}

To calculate the thermodynamic properties of our clusters we use
the superposition method;\cite{Wales93a,JD95a} 
in this approach the density of states, $\Omega(E)$, is constructed using information about 
a set of minima on the PES (such as those represented in Figures \ref{38.minE} and \ref{55.minE}).
This choice is motivated by two considerations.
Firstly, methods such as multi-histogram Monte Carlo,\cite{Ferrenberg88,Labastie} 
often the method of choice for clusters, 
are not able to describe the low temperature thermodynamics for \LJ{38} correctly 
because in simulations the cluster is not able to cross the large free energy barrier 
between the fcc and icosahedral funnels. 
It might be possible to overcome this difficulty by using more advanced techniques such as 
jump-walking\cite{Frantz90} or umbrella sampling,\cite{Torrie77} 
however these methods can be computationally demanding. 
The advantage of the superposition method is that it can be used to calculate {\em absolute}
densities of states for different regions of phase space, and therefore there is no need
to cross any free energy barrier.
Secondly, the superposition method can give greater physical insight into the relationship
between the topography of the PES and the thermodynamics. 

We review the superposition method here because it will be important later when we formulate
the thermodynamic properties of a transformed energy surface, and
because some additional developments need to be made before it can be applied to \LJ{38}.
The fundamental relation is
\begin{equation}
\Omega(E)=\sum_{E_s<E}n_s\Omega_s(E), 
\label{equation:supern}
\end{equation}
where 
the sum is over {\em all} the geometrically distinct minima on the PES,
$\Omega_s(E)$ is the density of states of a single 
minimum $s$, $E_s$ is the potential energy of minimum $s$, 
and $n_s$, the number of permutational isomers of minimum $s$, is given by $n_s=2N!/h_s$ where
$h_s$ is the order of the point group of $s$. 
This equation is exact and just expresses the division of configuration space into
basins of attraction that surround each minimum on the PES.\cite{StillW82}
One difficulty with the above equation is that $\Omega_s(E)$ is not known {\it a priori\/}.
However, if the basin is assumed to be harmonic then the form of $\Omega_s(E)$ is simple, 
and using this expression a qualitative picture of the thermodynamics can be obtained.\cite{Wales93a}
Furthermore, anharmonic expressions for $\Omega_s(E)$ have also been derived which are able to 
describe the thermodynamics of Lennard-Jones clusters more accurately.\cite{JD95a}

The second difficulty with equation (\ref{equation:supern}) is that for all but the very smallest
clusters the sum involves an impractically large number of minima.
Hoare and McInnes,\cite{HoareM76} and more recently Tsai and Jordan,\cite{Tsai93a} have enumerated
lower bounds to the number of geometric isomers for LJ clusters from 6 to 13 atoms.
This number rises exponentially with $N$.
Extrapolating the trend gives for LJ$_{55}$ an estimate of $10^{21}$ geometric isomers.
In such a case, as it is not possible to obtain a complete set of minima, one instead has to use a
representative sample, which, for example, 
could be obtained by performing  minimizations from a large set of configurations generated 
by a molecular dynamics trajectory.
However, one needs a way to correct for the incomplete nature of the sample.
This correction can be achieved by weighting the density of states for each {\it known} minimum
by $g_s$, the number of minima of energy $\sim E_s$ for which the minimum $s$ is representative.
Hence,
\begin{equation}
\Omega(E)\approx\sum_{E_{s}<E}' g_{s} n_{s}\Omega_{s}(E)
\end{equation}
where the prime indicates that the sum is now over an {\it incomplete} but {\it representative} sample of minima.
The effect of $g_s$ can be incorporated by using $p_s(E')$, the probability of obtaining $s$
in a minimization from a configuration generated by a microcanonical simulation
at energy $E'$. 
If the system is ergodic on the time scale of the simulation,
then $p_s(E')\propto g_s n_s \Omega(E')_s$.
Hence,
\begin{equation}
\Omega(E)\propto\sum_{E_{s}<E}' p_s \left(E' \right) {\Omega_s(E)\over\Omega_s(E')}.
\label{equation:superp}
\end{equation}
Since we know the low energy limiting form of $\Omega(E)$, where the contribution of the lowest energy minimum 
dominates, the proportionality constant in (4) can also be obtained.

This reweighting technique is analogous to histogram Monte Carlo,\cite{Ferrenberg88,McDonald,Ferrenberg89}
but instead of determining the configurational density of states from the canonical potential energy
distribution, $g$, effectively a density of minima, is found from the microcanonical probability
distribution of being in the different basins of attraction. Hence there are some similarities
with the microcanonical multihistogram approach of Calvo and Labastie.\cite{Calvo95}

The accuracy of the method depends on the statistical accuracy of $p_s(E')$; 
this probability distribution needs to be obtained at an energy where all relevant minima
are significantly sampled. This is possible for \LJ{55};\cite{JD95a} however, for \LJ{38} there is no
energy at which both the fcc and icosahedral minima can both be sampled because of the large 
free energy barrier. Therefore, in the latter case the density of states is obtained 
by adding two terms:
\begin{equation}
\Omega(E)=\Omega_{\rm fcc}(E)+\Omega_{\rm rest}(E).
\end{equation}
The density of states of the fcc region of phase space, $\Omega_{\rm fcc}(E)$, can be obtained by
summing the density of states for all the low energy fcc minima using 
equation (\ref{equation:supern}). (In fact the summation can be dispensed with since
the term from the global minimum is the only term in $\Omega_{\rm fcc}(E)$ that ever contributes significantly to
$\Omega(E)$.) At the melting point an \LJ{38} cluster samples both the icosahedral and the liquid-like 
regions of configuration space and so equation (\ref{equation:superp}) can be 
used to find $\Omega_{\rm rest}(E)$.
Once $\Omega(E)$ has been determined, various thermodynamic quantities can be calculated by the 
application of standard thermodynamical formulae.\cite{JD95a}

\begin{figure}
\begin{center}
\epsfig{figure=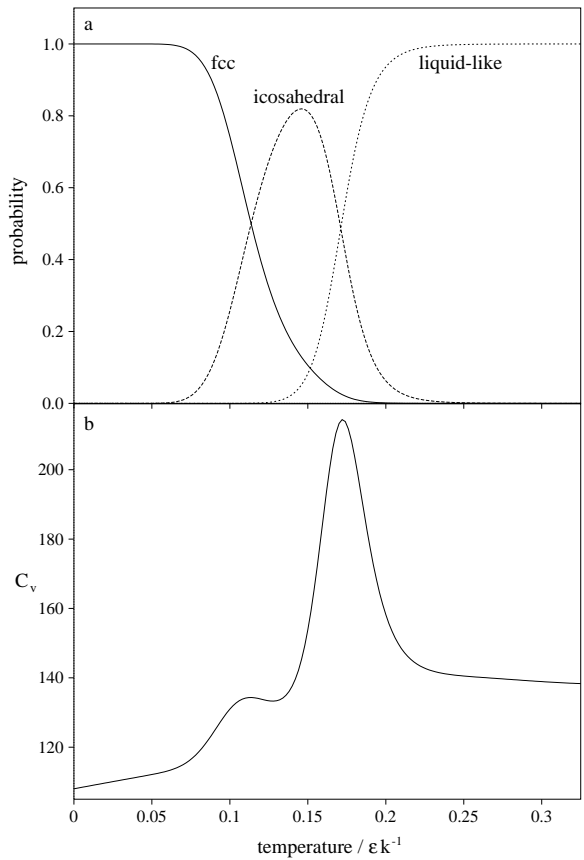,width=8.2cm}
\vglue0.1cm
\begin{minipage}{8.5cm}
\caption{\label{38un.therm} Equilibrium thermodynamic properties of the
untransformed LJ$_{38}$ PES in the canonical ensemble.
(a) The probability of the cluster being in the fcc, icosahedral
and `liquid-like' regions of bound configuration space.
(b) The heat capacity, $C_v$.
The sample of 661 minima was obtained from 2500 quenches along a microcanonical MD run
at $E=-150.1\epsilon$.}
\end{minipage}
\end{center}
\end{figure}

The resulting equilibrium thermodynamic properties of \LJ{38} 
are shown in Figure \ref{38un.therm}.
The anharmonic expression for $\Omega_s$ was used with 
the anharmonicity parameters\cite{anharm} determined by the technique give in ref. \onlinecite{JD95a}.
As there are many low energy icosahedral minima and as these minima have a lower mean
vibrational frequency than the truncated octahedron, 
the configurational entropy associated with the icosahedral funnel is 
larger than for the fcc funnel. 
As a result of this entropy difference there is a transition from the fcc to the 
icosahedral regions of configurational space at low temperature.
The transition is centred at a temperature of $\sim 0.11\,\epsilon k^{-1}$
and gives rise to the small peak in the heat capacity (Figure \ref{38un.therm}b).
The finite-size analogue of the bulk melting transition occurs
at $\sim 0.17\,\epsilon k^{-1}$, producing the main peak in the heat capacity.\cite{Calvoloop}
These transitions are also reflected in the occupation probabilities for the
different `phases' (Figure \ref{38un.therm}a).

We can understand the effect these thermodynamic transitions have on
global optimization by considering the result of cooling a cluster 
from the liquid-like state. The cluster is thermodynamically 
much more likely to enter the icosahedral funnel on quenching
because the free energy associated with the icosahedral structures at the melting
point is lower than that associated with the fcc structures. 
Furthermore, this effect is reinforced by the topology of the PES. 
The structure of simple atomic liquids has significant polytetrahedral
character;\cite{NelsonS,JD96b} 
in contrast fcc structures have no polytetrahedral character, and icosahedra are somewhere in between.
As a result of this greater structural similarity, 
the icosahedral funnel is more accessible from the liquid-like state.
(A similar effect has been noted by Straley who found that crystallization from 
a simple liquid is much easier in a curved space, where a polytetrahedral 
packing is the global minimum, than in normal flat space where a close-packed 
crystal is most stable.\cite{Straley84,Straley86})
The combined effect of these two features is to make it extremely probable
that on cooling the cluster will enter the icosahedral funnel.
On further cooling the cluster will remain trapped in this funnel
even when the fcc structures become lower in free energy because of the 
large free energy barrier between the two funnels. 
Clearly this behavior will present difficulties for 
annealing methods, which simply try to follow the free energy global minimum
down to zero temperature by gradual cooling.

The only opportunity to find the truncated octahedron by conventional simulation
is at high temperature in the narrow window where 
the fcc structures have a small, but not yet insignificant, 
probability of being occupied and where the Boltzmann weights for the intermediate states,
which mediate transitions between the fcc and icosahedral funnels, are large enough to 
make the free energy barrier surmountable.
Indeed, we did observe the truncated octahedron in the microcanonical simulation ($T=0.185\epsilon k^{-1}$) 
from which we obtained the distribution of minima depicted in Figure \ref{38.minE}.
However, minimization only led to this structure {\it once} in the entire
0.25$\,\mu$s run.\cite{conversion}

\begin{figure}
\begin{center}
\epsfig{figure=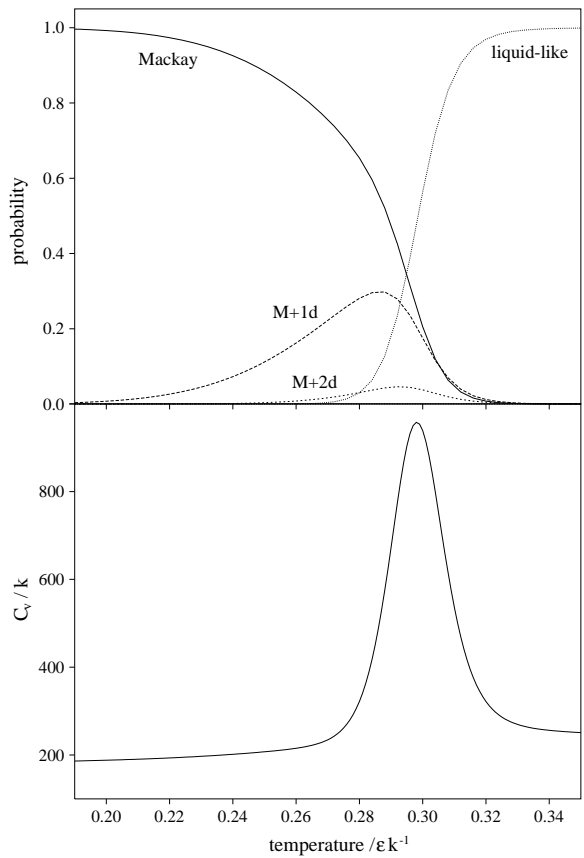,width=8.2cm}
\vglue0.1cm
\begin{minipage}{8.5cm}
\caption{\label{55un.therm} Equilibrium thermodynamic properties of the
untransformed LJ$_{55}$ PES in the canonical ensemble.
(a) The probability of the cluster being in the Mackay icosahedron, with
one or two defects, and `liquid-like' regions of bound configuration space.
(b) The heat capacity, $C_v$.
}
\end{minipage}
\end{center}
\end{figure}

Some thermodynamic properties of \LJ{55} are shown in Figure \ref{55un.therm}. 
The partition function for this system was obtained previously in the development
of the superposition method,\cite{JD95a} and it gives very good agreement with 
thermodynamic properties derived by more conventional means.\cite{JD95a,PhD3}
There is a single peak in the heat capacity curve
corresponding to the melting transition at which the cluster passes from the 
Mackay icosahedron (perhaps with one or two defects) into the liquid-like state.
For simulations in the middle of this melting region the cluster passes back and forth 
between the different states.\cite{Kunz93,Kunz94}
Similarly, on cooling from the liquid the cluster 
enters the icosahedral funnel relatively easily.
\LJ{55}'s single funnel PES presents no thermodynamic obstacles to global optimization.

\section{Global optimization by basin-hopping}

The global optimization method that we analyze here belongs to the `hypersurface deformation' 
family of methods.\cite{StillW88}
In this approach the potential energy function is transformed in a way that is hoped to make 
global optimization easier.
The transformations usually try to lower the number of local minima---thus 
reducing the search space---and/or the barrier heights between minima---thus
increasing the rates of transitions between minima. 
However, little attention is usually paid to the effects of the transformation on the 
thermodynamics of the system. 

Once the global minimum of the transformed PES is found it is mapped back to the 
original surface in the {\it hope} that this will lead to the global minimum on 
the original PES. 
However, there is no guarantee that the global minima on the two surfaces
are related and often there are good reasons to think that they are not.
For example, one method suggested for clusters is to increase the range of the potential.\cite{StillD90b,Shao98}
Such a transformation can dramatically reduce the number of minima\cite{JD96b,StillD90b}
and also lower the barrier heights.\cite{Wales94b,PhD6,Miller98}
However, it has been shown that changing the range of the potential for clusters can lead to many 
changes in the identity of the global minimum.\cite{JD95c,Chang,JD97e}

In contrast, the transformation that we apply to the PES is guaranteed 
to preserve the identity of the global minimum.
The transformed potential energy $\tilde E_c$ is defined by:
\begin{equation}
 \tilde E_c({\bf X}) = {\rm min}\left\{ E_c({\bf X}) \right\},
\end{equation}
where ${\bf X}$ represents the vector of nuclear coordinates
and min signifies that an energy minimization is performed starting
from ${\bf X}$.
Hence the energy at any point in configuration space is assigned to that of the local
minimum obtained by the minimization, and the transformed PES consists of
a set of plateaus or steps each corresponding to the basin of attraction surrounding a minimum on
the original PES.

\begin{figure}
\begin{center}
\epsfig{figure=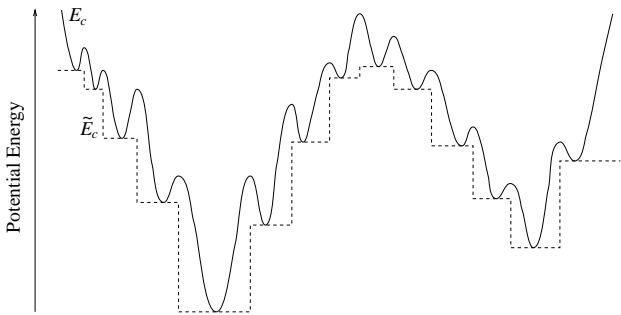,width=8.2cm}
\vglue 0.2cm
\begin{minipage}{8.5cm}
\caption{\label{fig:stair}
A schematic diagram illustrating the effects of our potential
energy transformation for a one-dimensional example.
The solid line is the potential energy of the original surface and the dashed line is the
transformed energy, $\tilde E_c$.}
\end{minipage}
\end{center}
\end{figure}

One can easily see that the transformation will have a significant effect on the dynamics.
Transitions to a lower energy step are barrierless, and so relaxation down a funnel is like
descending a multi-dimensional staircase (Figure \ref{fig:stair}).
Furthermore transitions can occur at any point along the boundary between basins of attraction,
whereas transitions on the untransformed surface can occur only when the system
passes along the transition state valley.
As a result intrawell vibrational motion is removed and the system can hop
directly between basins of attraction at each simulation step, hence our name for the 
algorithm, `basin-hopping'. 
In structural terms the system can now pass through large repulsive barriers on the 
untransformed PES.
Similarly, if the method were applied to a polymer system it should remove entanglement 
effects because chains can pass through each other.

However, the transformation does not entirely remove the energy barriers between two funnels but only 
the component due to transition states (Figure \ref{fig:stair}).
Therefore, it is not self-evident that it will aid global optimization on
multiple-funnel PES's; this effect depends on how the transformation affects the thermodynamics
of the system.

To explore the transformed PES we simply use a canonical Monte Carlo simulation.
For clusters we also have to consider how to restrict our search of configuration space 
to that for bound clusters. 
This problem is more pressing than for the original PES, since the transformation has 
removed many of the energy barriers to the dissociation of the cluster.
We have considered two approaches to achieve this. 
The first is to place the cluster in a tight-fitting box;
the second is to reset the coordinates at each successful step to the relevant minimum.
In the latter case the method is equivalent to Li and Scheraga's Monte Carlo 
Minimization approach,\cite{Li87a} although they did not conceive of the method in terms
of a transformation of the PES.
White and Mayne have now shown that resetting the coordinates is probably the best
strategy,\cite{whitem98} and it is the one we concentrate on here.

We have found that the `basin-hopping' algorithm performs well. 
In our application to Lennard-Jones clusters it found all the known lowest
energy configurations up to 110 atoms, including those with 
non-icosahedral global minima.\cite{WalesD97}
Furthermore, it has performed impressively for a wide range of cluster systems
some of which have a considerably more rugged PES than Lennard-Jones clusters.\cite{JD97e,JD98b,WalesH98} 

In Figure \ref{38gmin.time} we show some basin-hopping trajectories for \LJ{38} and \LJ{55}. 
It can be seen that the cluster is able to pass between the fcc and icosahedral funnels of \LJ{38}
over a range of temperature.
Therefore, the free energy barriers must be lowered by the transformation.
In the next section we will examine the thermodynamics of the transformed PES
to discover the basis for this change. 
The predicted acceleration of the dynamics is also evident from the frequency of 
transitions between the different states of \LJ{38} and \LJ{55} (Figure \ref{38gmin.time}). 
The timescale (in numbers of steps) for these processes is much more 
rapid than for similar processes on the original PES, where even diffusion across 
a flat free energy landscape can be slow. 
However, it should be remembered that a single MC step on $\tilde E_c$ is much more computationally expensive than an
MD step or an MC cycle on the untransformed PES, because it involves a minimization.

\subsection{Thermodynamics for $\tilde E_c$}

\begin{figure}
\begin{center}
\epsfig{figure=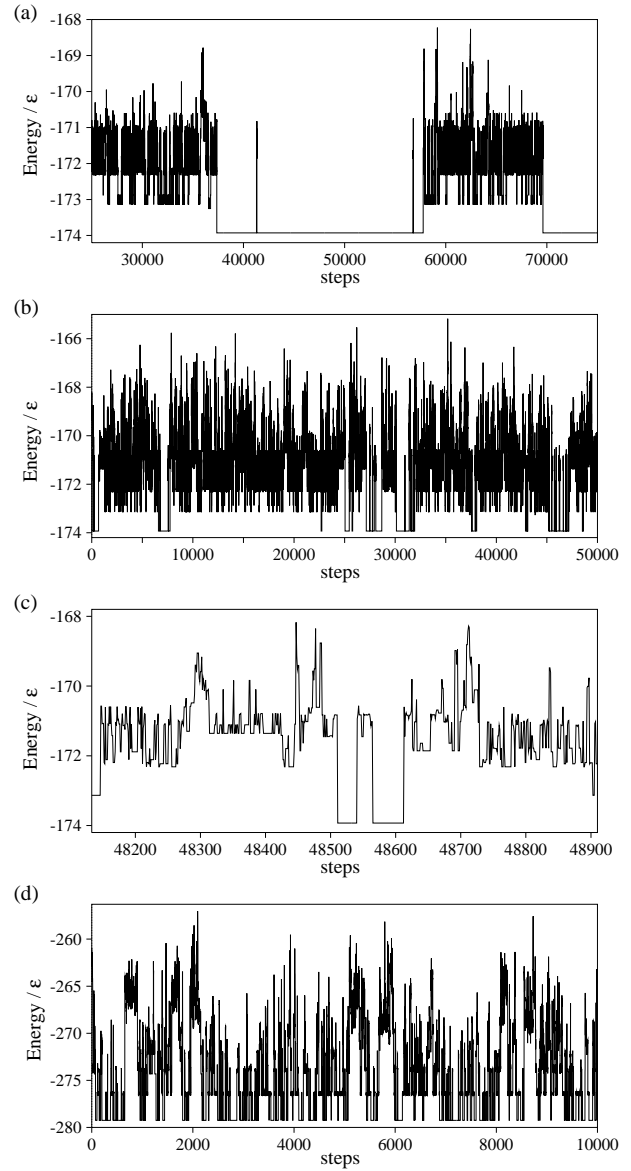,width=8.2cm}
\vglue0.1cm
\begin{minipage}{8.5cm}
\caption{\label{38gmin.time} 
$\tilde E_c$ as a function of the number of steps in 
basin-hopping runs for \LJ{38} at (a) $T=0.4\epsilon k^{-1}$ and (b) $T=0.9\epsilon k^{-1}$
In (c) we show a close-up of one section of (b) in which the system passes from
the icosahedral region of configuration space to the fcc region and then back again.
(d) A basin-hopping run at $T=10.0\epsilon k^{-1}$ for LJ$_{55}$.
}
\end{minipage}
\end{center}
\end{figure}

To calculate the thermodynamics for the transformed energy surface, $\tilde E_c$, 
the partition of configuration space that we used in section \ref{sect:thermo.un}
(equation (\ref{equation:supern})) is appropriate since
the plateaus of $\tilde E_c$ correspond to the basins of attraction on
the original PES. However, we need to derive the form of $\Omega_s(E)$. 
The configurational density of states of a plateau $s$, $\Omega_{c,s}(\tilde E_c)$ 
is simply 
\begin{equation}
\Omega_{c,s}(\tilde E_c)=\delta(\tilde E_c-E_s) A_s
\end{equation}
where $A_s$ is the hyperarea of the basin of attraction of minimum $s$. 
This expression is then convoluted with the kinetic density of states
to give 
\begin{equation}
\Omega_s(E)={(2\pi m)^{\kappa/2}\over \Gamma(\kappa/2) h^\kappa} A_s (E-E_s)^{\kappa/2-1} \theta(E-E_s),
\end{equation}
where $m$ is the mass of an atom, $h$ is Planck's constant, 
$\kappa$, the number of vibrational degrees of freedom, is $3N-6$ and $\theta$ is the Heaviside step function.
As $A_s$ cannot be readily calculated, we cannot use equation (\ref{equation:supern}) 
to obtain the total density of states but instead 
employ equation (\ref{equation:superp}). 
For $\tilde E_c$ the latter approach can even be applied to \LJ{38} since, 
as Figure \ref{38gmin.time}b shows, conditions can
be found where the fcc, icosahedral and liquid-like regions of configurational
space are all significantly sampled. However, since we do our simulations on $\tilde E_c$
using standard Metropolis Monte Carlo sampling, we actually use the 
canonical analogue of equation (\ref{equation:superp}): 
\begin{equation}
Z(\beta)\propto \sum_s' p_s(\beta') {Z_s(\beta)\over Z_s(\beta')}, 
\label{equation:zsuperp} 
\end{equation}
where $Z$ is the partition function, the inverse temperature $\beta=1/kT$
and the probability distribution was obtained from a basin-hopping trajectory at $\beta'$.
To use this equation we must first Laplace transform the expression for $\Omega_s$
in (8) to give $Z_s$, the partition function of a plateau:
\begin{equation}
Z_s(\beta)=\left({2\pi m\over \beta h^2}\right)^{\kappa/2} A_s e^{-\beta E_s}.
\end{equation}
Equation (\ref{equation:zsuperp}) then reduces to
\begin{equation}
Z(\beta)\propto \sum_s' p_s(\beta') e^{-E_s(\beta-\beta')}.
\label{eq:Ztrans}
\end{equation}
It is interesting to note that this equation is identical to that used in histogram 
MC.\cite{Ferrenberg88,McDonald,Ferrenberg89}
This equivalence stems from the fact that the potential energy is constant on each plateau $s$.
Hence the multi-histogram techniques, in which information from a number of runs 
at different temperatures is used to construct the partition function, could also be applied.
However, such an approach was not necessary in this study.

The $A_s$ values depend upon how configuration space is restricted to bound clusters.
When resetting the coordinates to those of the minimum at each successful step,
configuration space is restricted to a hypersphere (with radius corresponding to
the maximum step size) around each local minimum, and so the thermodynamics 
(and the performance of the basin-hopping algorithm\cite{White98}) is
somewhat dependent on the maximum step size used.
When using a container the total accessible configuration space, $\sum n_s A_s$,
is the volume of a $\kappa$-dimensional hypersphere whose radius corresponds
to that of the container.
Therefore, the thermodynamic contribution of irrelevant regions of configuration
space corresponding to clusters of low density, or with atoms evaporated, increases
with container size.
Hence, it is advantageous to use a tight-fitting container, 
and for this situation the thermodynamics are similar to when 
the coordinates are reset.

One further consideration is that one can imagine situations where resetting
the coordinates breaks detailed balance.
For two minima A and B which are adjacent on the PES, it may 
be that no part of the basin of attraction of minimum B is within the 
maximum step size of minimum A, whereas the basin of attraction of A lies within the
maximum step size of minimum B. If so the system can pass from B to A but never from
A to B, thus breaking the detailed balance condition.
Although this means that the `resetting' implementation of the basin-hopping
algorithm might not formally produce a canonical ensemble, in practice it does 
to a good approximation---the thermodynamic properties
calculated using equation (\ref{eq:Ztrans}) for samples generated 
at different temperatures are in reasonable agreement.

\begin{figure}
\begin{center}
\epsfig{figure=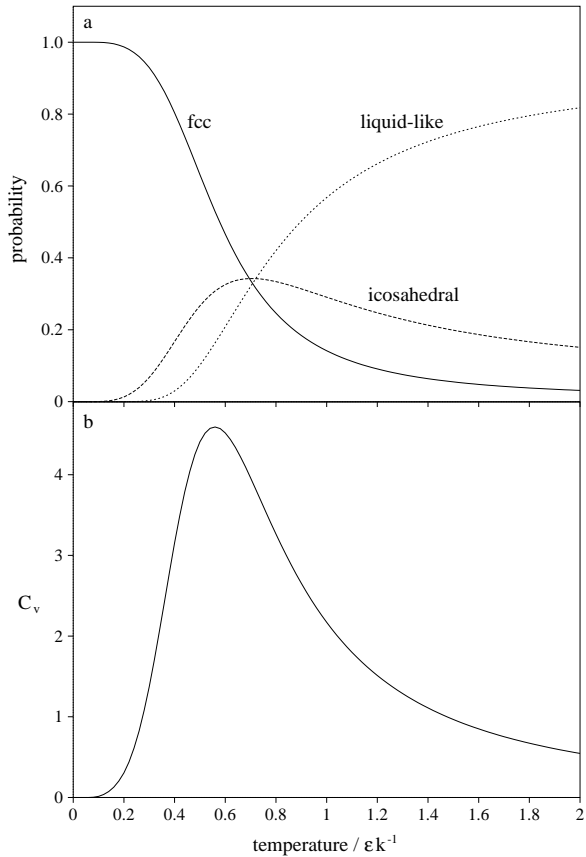,width=8.2cm}
\vglue0.1cm
\begin{minipage}{8.5cm}
\caption{\label{38trans.therm} Equilibrium thermodynamic properties of the transformed LJ$_{38}$ PES
when the maximum step size is fixed at 0.35$\sigma$.
(a) The probability of the cluster being in the fcc, icosahedral
and `liquid-like' regions of bound configuration space.
(b) The configurational component of the heat capacity.
These properties were Calculated from a probability distribution 
produced by a 500000 step run at $T=0.9\epsilon k^{-1}$.
}
\end{minipage}
\end{center}
\end{figure}

For the transformation of the LJ$_{38}$ PES to reduce the free energy barriers between the funnels,
the probability of the system occupying the intermediate states
between them must be non-negligible under conditions where the
icosahedral and fcc structures also have significant occupation probabilities.
The thermodynamics of the transformed PES have just these properties
(Figure \ref{38trans.therm}).\cite{prldif}
The transitions have been smeared out and there is now only one broad peak in the heat capacity.
Most significantly, the probability that the system is in the basin of
attraction of the global minimum decays much more slowly,
and the temperature range for which both the liquid-like minima
and the global minimum have significant probabilities is large.
Basin-hopping runs anywhere in this temperature range are able to locate
the global minimum rapidly from a random starting point. 
At the lower temperatures in this range, e.g. $T=0.4\epsilon k^{-1}$,
there is still a small free energy barrier and so the cluster 
can be trapped in one of the funnels for many
steps (Figure \ref{38gmin.time}a), however at higher temperatures 
the system passes more rapidly between the states (Figure \ref{38gmin.time}b and c).

\begin{figure}
\begin{center}
\epsfig{figure=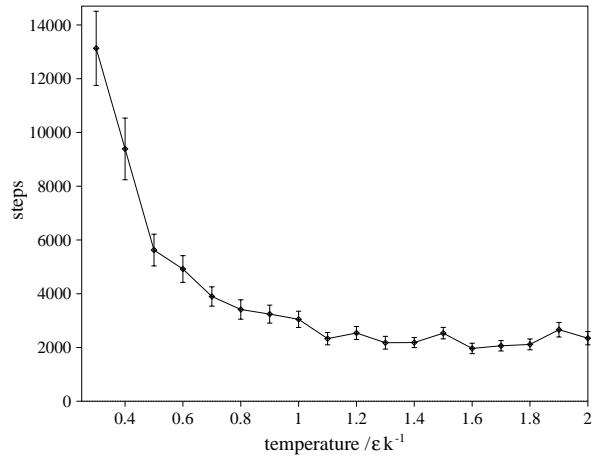,width=8.2cm}
\vglue0.1cm
\begin{minipage}{8.5cm}
\caption{\label{38.fpt} Mean first-passage time for 
the basin-hopping algorithm to reach the LJ$_{38}$ global minimum as a function of temperature
when the maximum step size is fixed at 0.35$\sigma$.
Each point represents an average over 100 runs starting from a random configuration. 
}
\end{minipage}
\end{center}
\end{figure}

In this way we can correlate the performance of the 
basin-hopping algorithm with the thermodynamics. 
The temperature dependence of the first-passage time for reaching the global minimum from a random configuration
is shown in Figure \ref{38.fpt}. 
As the temperature decreases the probability of the system being in the high energy 
intermediate states between the two systems decreases. 
The resulting increasing free energy barrier causes the first-passage 
time to rise at low temperatures. 
Interestingly, and perhaps unexpectedly, there is little rise at higher temperature.
For many systems, e.g.\ proteins,\cite{Gutin98,Socci96}
the first-passage time has a minimum as a function of temperature;
the rise at high temperature is because the equilibrium probability of 
being in the global minimum tends to zero.
However, for the basin-hopping algorithm the first-passage time at 
high temperature is approximately constant.\cite{acceptanceratio}  
In fact the probability of being in the global 
minimum never goes to zero, 
e.g.\ $p_{O_h}(T=\infty)=0.0024$. 

\begin{figure}
\begin{center}
\epsfig{figure=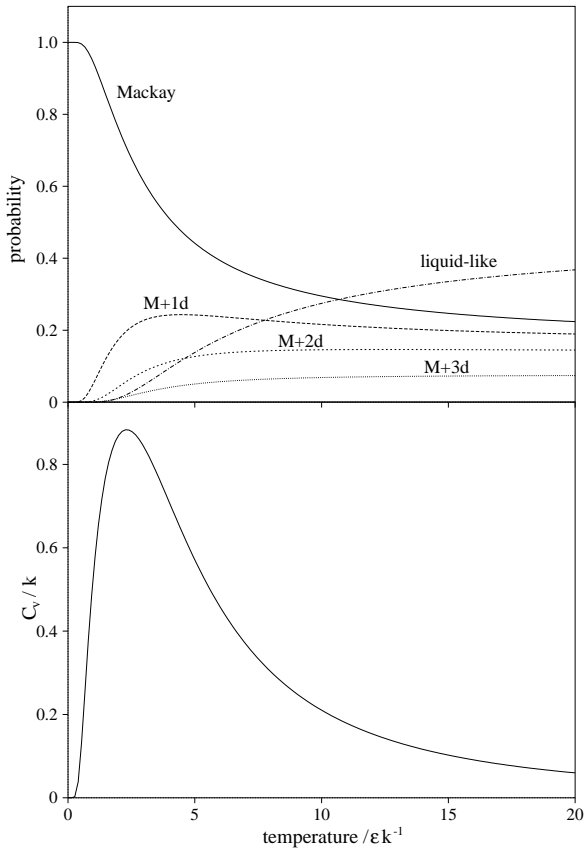,width=8.2cm}
\vglue0.1cm
\begin{minipage}{8.5cm}
\caption{\label{55trans.therm} Equilibrium thermodynamic properties of the transformed LJ$_{55}$ PES
when the maximum step size is fixed at 0.35$\sigma$.
(a) The probability of the cluster being in the Mackay icosahedron, with
one to three defects, and in `liquid-like' regions of bound configuration space.
(b) The configurational component of the heat capacity.
The properties were calculated from a probability distribution produced by a 
100000 step run at $T=10.0\epsilon k^{-1}$.
}
\end{minipage}
\end{center}
\end{figure}

\begin{figure}
\begin{center}
\epsfig{figure=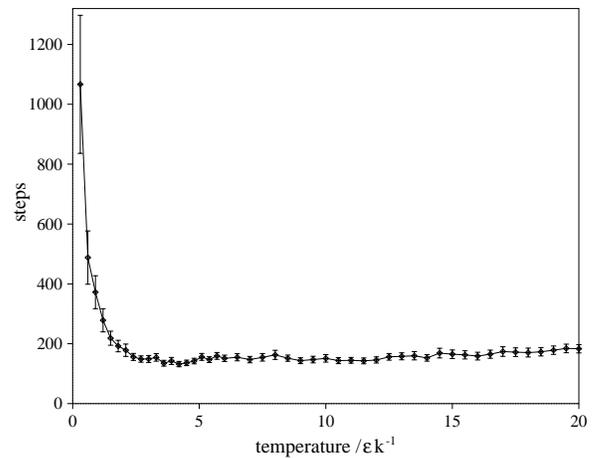,width=8.2cm}
\vglue0.1cm
\begin{minipage}{8.5cm}
\caption{\label{55.fpt} Temperature dependence of the mean first-passage time for 
the basin-hopping algorithm to reach the LJ$_{55}$ global minimum 
when the maximum step size is fixed at 0.35$\sigma$.
Each point represents an average over 100 runs starting from a random configuration.
}
\end{minipage}
\end{center}
\end{figure}

The thermodynamics on the transformed surface for \LJ{55} are also considerably 
broadened and the transitions have been smeared out over a large 
temperature range (Figure \ref{55trans.therm}).
Consistent with this, there is a greater probability that one of the states
intermediate between the global minimum and the liquid is occupied.
As expected for the single funnel topography of \LJ{55}, the basin-hopping 
algorithm needs far fewer steps to find the global minimum than for \LJ{38}.
It is remarkable that for a wide range of temperature ($T>1.5\epsilon k^{-1}$) 
the method on average requires fewer than 200 minimizations to find the global minimum 
from a random starting point (Figure \ref{55.fpt}) despite the estimated $10^{21}$ minima on the PES.
This is a testament not just to the efficiency of the basin-hopping algorithm 
but also to the dramatic effect that a funnel has in guiding the system towards the global minimum.
As with \LJ{38} the first-passage time is approximately constant at high temperature (Figure \ref{55.fpt});
$p_{I_h}(T=\infty)=0.11$. 

The superposition method allows us to connect the thermodynamics to the
properties of the PES. Hence we can understand the thermodynamics of the transformed surface
by examining expressions
for the canonical probability that the system is in a minimum or on a plateau $s$.
On the untransformed PES 
\begin{equation}
p_s(\beta)={n_s\exp(-\beta E_s) \over
                  \overline\nu_s^{\kappa}}\big/
             \sum_s{n_s\exp(-\beta E_s) \over
                   \overline\nu_s^{\kappa}},
\label{eq:ps.un}
\end{equation}
where $\overline\nu_s$ is the geometric mean vibrational frequency of minimum $s$,
and we have used the harmonic approximation---the more accurate anharmonic form\cite{JD95a} is 
complicated and does not provide any additional physical insight.
For the transformed PES
\begin{equation}
p_s(\beta)={n_s A_s\exp(-\beta E_s)\over
            \sum_s n_s A_s \exp(-\beta E_s)}.
\label{eq:ps.trans}
\end{equation}
In equations (\ref{eq:ps.un}) and (\ref{eq:ps.trans}) 
the sums are over {\it all} the minima on the potential energy surface.

\begin{figure}
\begin{center}
\epsfig{figure=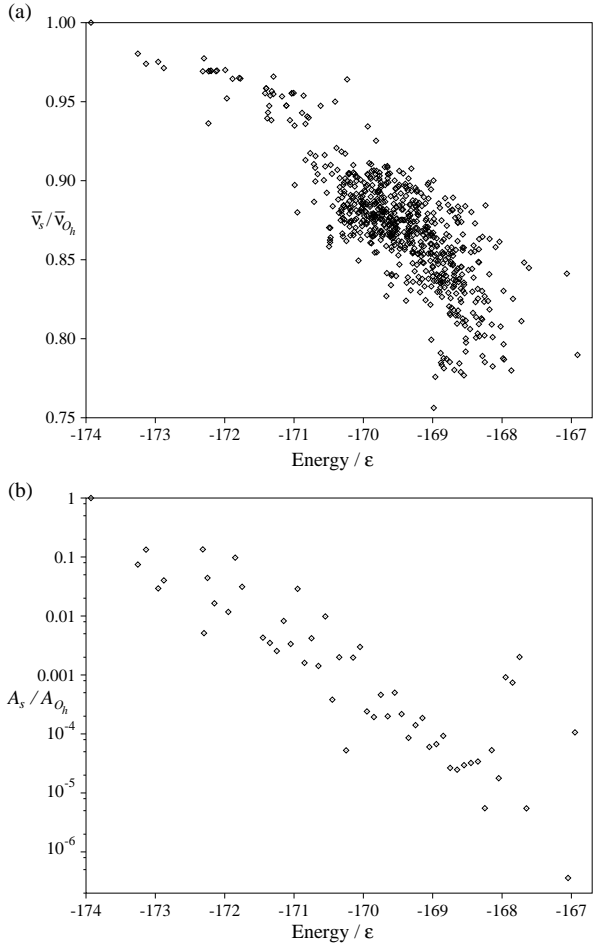,width=8.2cm}
\vglue0.1cm
\begin{minipage}{8.5cm}
\caption{\label{38.vibAs} 
(a) Geometric mean vibrational frequencies, $\overline\nu_s$, and
(b) basin of attraction areas, $A_s$, for a sample of \LJ{38} minima.
}
\end{minipage}
\end{center}
\end{figure}

The two expressions (\ref{eq:ps.un}) and (\ref{eq:ps.trans}) differ only in the vibrational frequency and $A_s$ terms.
The fcc to icosahedral and the icosahedral to liquid transitions are caused by
the greater number of minima (and therefore the larger entropy) associated with the 
higher temperature states.
On the untransformed surface this effect is reinforced by the decrease in the
mean vibrational frequencies with increasing potential energy
(Figure \ref{38.vibAs}a).
Although the dependence of $\overline\nu_s$ on $E_s$ may seem relatively small, 
because $\overline\nu_s$ is raised to the power $3N-6$ it leads
to a significant increase in the entropy of the higher energy states, sharpening
the transitions and lowering the temperature at which they occur.
In contrast, $A_s$ decreases rapidly with increasing potential energy,
(Figure \ref{38.vibAs}b).
The decrease in $A_s$ reduces the entropy of the higher energy states,
causing the transitions to be broadened and the temperature
at which they occur to increase.

To obtain the $A_s$ values shown in Figure \ref{38.vibAs}b we used $p_s(\beta')$, the probability with which 
a plateau $s$ was visited in a basin-hopping trajectory.
As $p_s(\beta)\propto g_s n_s A_s \exp(-\beta E_s)$, relative values of $g_s A_s$ can be found.
A similar inversion of $p_s(E')$ for the untransformed surface allows values of $g_s$ to be found.\cite{JD95a}
However, the $g_s$ and $g_s A_s$ values are for different sets of minima.
The two samples of minima overlap completely for the low energy minima, and so $A_s$ values for these
minima can be calculated directly.
For higher potential energies, where the spectrum of minima is quasi-continuous, we find 
the average value of $A_s$ in an interval of the potential energy, $\Delta E_c$, 
from $\sum_s g_s A_s/\sum_s g_s$ where the sums are only over the minima in the two samples 
which occur in that energy interval.

The superposition approach also allows us to comment on the thermodynamics of
a related optimization algorithm which steps directly between connected minima on
the PES (the steps are achieved using a transition state searching algorithm).\cite{JD97a,Barkema96a,Mousseau97}
As with the basin-hopping algorithm the detrimental effects of vibrational motion are thus removed.
This method has been successfully used to find low energy structures of amorphous 
semiconductors\cite{Barkema96a,Mousseau97,Mousseau97b} and 
its performance for a few Lennard-Jones clusters has also been illustrated.\cite{JD97a}
If each step is accepted with a Metropolis criterion then it is easy to formulate
the thermodynamics for this discrete space of minima:\cite{mmdetailed}
\begin{equation}
p_s(\beta)={n_s \exp(-\beta E_s)\over
            \sum_s n_s \exp(-\beta E_s)}.
\end{equation}
This expression is intermediate between equations (\ref{eq:ps.un}) and (\ref{eq:ps.trans}).
The thermodynamics have been broadened with respect to the original PES, 
but not as much as for the `basin-hopping' transformation of the PES. 
The above approach is probably less efficient than `basin-hopping' because 
of the expense of searching for a transition state at each step. 
However, it has the advantage that interesting information about the PES, such as
reaction pathways for complex processes\cite{JD97a,Mousseau97,JD98c,Barkema98a} 
and the connectivity patterns between the minima,\cite{WalesMW98} can also be obtained.

\section{Conclusions}

We can now explain in detail why
the basin-hopping or Monte Carlo Minimization\cite{Li87a} method is successful.
First, the staircase transformation removes the transition state barriers between minima
on the PES without changing the identity of the global minimum,
accelerating the dynamics.
Second, it changes the thermodynamics of the surface, broadening the transitions
so that even for a multiple-funnel surface such as that of \LJ{38},
the global minimum has a significant probability of occupation at temperatures
where the free energy barrier for passage between the funnels is surmountable.

It is this latter feature which is especially important in overcoming
the difficulties associated with multiple funnels and represents a
new criterion for designing successful global optimization methods.
Most previous hypersurface deformation schemes have been developed without 
regard to the thermodynamic effects of the transformation and 
so, in some cases, they may make optimization more difficult.
The use of Tsallis statistics\cite{penna95,tsalliss96,andricoaeis96,Andricioaei97,hansmann97}
to improve annealing algorithms is another example of how thermodynamic insight can be used.

The broadened transition also means that the global minimum can be
found by simulations over a relatively broad range of temperature.
Therefore, simple canonical Monte Carlo is sufficient to search the transformed surface
and the performance is relatively insensitive to the choice of temperature.
However, any of the more advanced techniques for searching PES's 
such as simulated annealing,\cite{KirkSA} genetic algorithms,\cite{Deaven95}
jump walking,\cite{Frantz90} Tsallis statistics,\cite{Andricioaei97} 
and simulated tempering,\cite{Marinari92} could be used to search the transformed PES.
As many of these methods have been designed to work on untransformed PES's
it is not obvious that they will improve performance, 
and preliminary investigations have not yet indicated that any of them
provides a significant advantage.
It is also interesting to note that the most successful applications of 
genetic algorithms to clusters refine each new configuration generated by mutation or
mating by minimization;\cite{Deaven96,Niesse96a} this minimization step has been shown to be crucial.\cite{White98}
These genetic algorithms are therefore searching the transformed PES, $\tilde E_c$,
and it is probably this feature
which is responsible for their success. In fact, we have succeeded in obtaining
comparable results for Lennard-Jones clusters with a modified genetic algorithm.\cite{markham}
The most efficient method is likely to be system-dependent and draw upon features
from a number of different approaches.

\LJ{75} could prove to be an interesting example to explore the use of other
techniques to search the transformed surface because its
global minimum, even with the basin-hopping algorithm, is difficult to find. 
The free energy barriers between the decahedral and
fcc funnels, although reduced by the transformation of the PES, are still 
large enough to make global optimization difficult.
Therefore, \LJ{75} may provide a suitable system in which to investigate algorithms that enhance barrier 
crossing or simulate an ensemble which has broader probability distributions than 
the canonical ensemble.

The broadened transition that results from the staircase transformation also differs
markedly from the optimum conditions for protein folding,
if we assume that proteins have evolved single-funnel surfaces
in order to fold efficiently.
A steep funnel provides a large thermodynamic driving force
for relaxation to the global minimum\cite{JD96c,Bryngel95},
and also leads to a sharp thermodynamic transition.
However, in global optimizations one cannot make assumptions about the
topography of the PES, and on a multiple-funnel PES features such as
steepness can exacerbate the difficulties associated with trapping in
secondary funnels\cite{JD96c}.

Moreover, in protein folding there is the extra requirement that the folded
protein must remain localized in the native state, and this necessitates a
sharp transition.
There is no need for a global optimization method to mimic this
property. Indeed, when applied to a PES with multiple funnels
the optimum temperature for the basin-hopping approach
is above the transition, where the system only spends a minority of
its time associated with the global potential energy minimum.

\acknowledgements

D.J.W.\ is grateful to the Royal Society 
and M.A.M.\ to the Engineering and Physical Sciences Research Council 
for financial support.
The work of the FOM Institute is part of the
scientific program of FOM and is supported by the Nederlandse
Organisatie voor Wetenschappelijk Onderzoek (NWO).

\end{multicols}

\begin{thebibliography}{10}
\bibitem{pathnote}
The pathway shown here is considerably lower in energy than the one reported in
  Ref.\ \onlinecite{JD97a} and was the result of a more extensive search in
  which over 7000 minima and 10000 transition states were found.
\bibitem{anharm}
The values of the anharmonicity parameters were: $a_{O_h}$=0.39,
  $a_{icos}$=0.43 and $a_{liquid}$=0.40. The shape of the heat capacity curve
  and the position of the transitions are slightly sensitive to the values
  used. These values are an improvement on those used in ref.\
  \onlinecite{JD98a}.
\bibitem{Calvoloop}
It should be noted that the melting transition does not give rise to a van der
  Waals loop in the microcanonical caloric curve. The presence of a loop in the
  caloric curve obtained by Calvo using a multi-histogram Monte Carlo
  approach\cite{Calvo98} probably results from the cluster being trapped in the
  fcc funnel up until melting.
\bibitem{conversion}
The conversion of the time from reduced units, $(m\sigma^2/\epsilon)^{1/2}$,
  into seconds is based on the values of $\epsilon$, $m$ and $\sigma$
  appropriate for argon.
\bibitem{prldif}
The equilibrium thermodynamic properties of the transformed \LJ{38} and \LJ{55}
  PES's reported here are slightly different from ref.\ \onlinecite{JD98a}
  because of the dependence of the thermodynamics on the maximum step size. In
  ref.\ \onlinecite{JD98a} the average maximum step size was 0.38$\sigma$ for
  \LJ{38} and 0.41$\sigma$ for \LJ{55}.
\bibitem{acceptanceratio}
The temperature dependence of the first-passage time is considerably different
  if the maximum step size is dynamically adjusted to give a constant
  acceptance ratio. With this approach the first-passage time has a minimum,
  because the maximum step size increases at higher temperature, changing the
  thermodynamics and decreasing the equilibrium probability of being in the
  global minimum.
\bibitem{mmdetailed}
However, depending on the algorithm used to locate transition states, detailed
  balance may not be obeyed, and so the method may not formally simulate a
  canonical ensemble in the space of minima. In our experience with using the
  eigenvector-following technique to locate transition states, instances have
  been found where when we start from minimum A we can find a transition state
  that connects it to minimum B, but when we start from minimum B this
  transition state is never found, thus breaking detailed balance.
\end{thebibliography}
\end{document}